\documentclass[11pt]{article} 
\usepackage{graphicx,amssymb,mathrsfs,array,multirow}  
\begin{document}  

\vskip 20pt  
\renewcommand{\thefootnote}{\fnsymbol{footnote}}  

\begin{center}  
{\Large{\bf Relating small neutrino masses and mixing\footnote{Talk
given by A. Raychaudhuri at the International Conference on
Massive Neutrinos, IAS, NTU, Singapore, February 2015.} }}\\
\vspace*{1cm}  
{ {\sf Soumita Pramanick\footnote{e-mail: soumitapramanick5@gmail.com},} 
{\sf Amitava Raychaudhuri\footnote{e-mail: palitprof@gmail.com}}
} \\  
\vspace{10pt}  
{\small  {\em Department of Physics, University of Calcutta,  
92 Acharya Prafulla Chandra Road, Kolkata 700009, India}}
\normalsize

\end{center}  


\begin{abstract}

Experiments on neutrino oscillations have uncovered several small
parameters, $\theta_{13}$ being a prominent one. Others are the
solar mass splitting {\em vis-\`{a}-vis} the atmospheric one and
the deviation of $\theta_{23}$ from maximal mixing. In this talk
we elaborate on a  neutrino mass model based on the see-saw
mechanism in which  the mixing angles to start with are either
vanishing ($\theta_{13}$ and $\theta_{12}$) or $\pi/4$
($\theta_{23}$).  The atmospheric mass splitting is taken as
a part of this initial structure but the solar splitting is
absent.  A perturbative contribution, originating from a Type-I
see-saw,  results in  non-zero values of $\theta_{13}$,
$\theta_{12}$, $\Delta m^2_{solar}$,   and shifts $\theta_{23}$
slightly  from $\pi/4$, interrelating them all. The model
incorporates CP-violation, the phase $\delta$ being close to
3$\pi$/2  for (a) quasi-degeneracy or (b) inverted mass ordering.
It will be put to test as the neutrino parameters get
better determined.
\end{abstract}


\texttt{Key Words:~~Neutrino mixing, $\theta_{13}$, Leptonic CP-violation,
Neutrino Mass ordering, Perturbation}

\renewcommand{\thesection}{\Roman{section}} 
\setcounter{footnote}{0} 
\renewcommand{\thefootnote}{\arabic{footnote}} 
\noindent

\vskip 15pt

\section{Introduction}
Neutrino mass and mixing have been subjects of intensive
exploration as they shed light on the physics beyond the
standard model. Atmospheric and solar neutrinos indicate two very
different scales of neutrino mass splitting -- $\Delta
m^2_{solar}/|\Delta m^2_{atm}| \sim 10^{-2}$ -- which are
confirmed in accelerator and reactor  experiments. The lepton
mixing is captured in the PMNS matrix\footnote{We use the PDG
\cite{PDG} parametrization of the PMNS matrix.}. The global fits
to the data \cite{Gonzalez, Valle} from atmospheric, solar,
accelerator, and reactor experiments indicate $\theta_{13}$ to be
small \cite{t13} ($\sin \theta_{13} \sim 0.1$) and $\theta_{23}$
to be near maximal ($\sim \pi/4$).   Here we discuss a model in which
the atmospheric mass splitting with maximal mixing in this
sector, $\theta_{23} = \pi/4$, follows from  a zero-order mass
matrix which sets the scale of the problem. There is at this stage
no solar splitting and the other two mixing angles are also
absent\footnote{One mixing angle being 
$\pi/4$ and the other two zero can be a 
manifestation of some underlying symmetry.}.   
$\theta_{13}$ and a small shift to $\theta_{23}$  arise
from a Type-I see-saw \cite{seesaw} contribution which also
results in the solar mass splitting, acting as a perturbation. A
non-zero $\theta_{12}$ is also produced and due to the degeneracy
of masses it is not small. The three non-zero mixing angles open
the possibility of CP-violation in the lepton sector. This model
accommodates a CP-phase $\delta$ which must be close to maximal
($\delta \sim \pi/2, ~3\pi/2$) if the neutrinos have an inverted
mass ordering or if they are quasidegenerate \cite{spar14}.  Earlier work
which partially address similar issues can be traced to
\cite{old, pert}, but to our knowledge this is the first time
that {\em all} the small parameters  have been shown to have the
same perturbative origin  and are consistent with the latest
data.

\section{The model}
The starting choice of the mixing angles imply
the following form of the mixing matrix, the columns of it 
being the unperturbed flavour basis\footnote{In this flavour 
basis the charged lepton mass
matrix is taken to be diagonal.}:
\begin{equation}
U^0= \pmatrix{1 & 0 & 0 \cr
0 & \sqrt{1 \over 2} & \sqrt{1 \over 2} \cr
0 & -\sqrt{1 \over 2} & \sqrt{1 \over 2}} \;\;.
\label{mix0}
\end{equation}
Solar splitting is absent at this stage causing the 
first two mass eigenvalues to be degenerate\footnote{Due 
to this degeneracy $\theta_{12}$ is arbitrary
and can be chosen to be zero as done here.}. Thus the 
unperturbed neutrino mass matrix is 
$M^0 = {\rm diag} \{ m^{(0)}_1, m^{(0)}_1, m^{(0)}_3 \}$ in the
mass basis.
The unperturbed mass eigenvalues are made real and positive by 
suitable choices of the Majorana phases.
The atmospheric splitting is given by 
$\Delta m^2_{atm} = (m^{(0)}_3)^2 - (m^{(0)}_1)^2$.
It is useful to define $m^{\pm} = m_3^{(0)} \pm m_1^{(0)}$
and express the unperturbed mass matrix in flavour basis 
as,
\begin{equation}
(M^{0})^{flavour}= U^{0}
M^{0} U^{0T}= {1 \over 2}
\pmatrix{2m^{(0)}_1 & 0 & 0 \cr 0 & m^+ & m^- \cr 0 & m^- & m^+} 
\;\;.
\label{mflav0}
\end{equation}
As already hinted, the perturbation can originate from a Type-I
see-saw.  In order to reduce the number of independent parameters
the Dirac mass term is taken to be proportional to the identity,
i.e., $ M_D = m_D ~\mathbb{I}$, in the flavour basis.  This
choice completely determines the right-handed flavour
basis although the form of $M_R^{flavour}$ can be chosen at will
to suit our purpose. In the interest of minimality we seek
symmetric matrices with the fewest non-zero entries.  Five
texture zero matrices fail the invertibility
criterion\footnote{Existence of the inverse of $M_R$ is an
essential condition for the see-saw mechanism.} and therefore are
not pursued. Next we try four texture zero options.  By
examining the different alternatives it can be seen  that all the
perturbation goals that we have set for ourselves could be
achieved by only two such candidates out of which one is scripted
below\footnote{The other alternative is a mere
$2\leftrightarrow3$ exchange of this configuration and the
corresponding results vary only up to a relative sign.}:
\begin{equation}
 M_R^{flavour} ={m_R} 
\pmatrix{0 & x e^{-i\phi_1} & 0 \cr x e^{-i\phi_1} & 0 & 0 \cr 0 & 0
& y e^{-i\phi_2}}  \  \ ,
\label{mflav1}
\end{equation}
where $x, y$ are dimensionless constants of ${\cal O}(1)$.
The Dirac mass is kept real without any loss of generality.

\section{Real $M_R$ ($\phi_1 = 0
~{\rm or} ~\pi, \phi_2 = 0 ~{\rm or} ~\pi$)}\label{sec3}

For notational simplicity in this section the phases are not
written explicitly, instead $x$ ($y$) is taken as positive or
negative depending on whether $\phi_1$ ($\phi_2$) is 0 or $\pi$.

Employing Type-I see-saw one can write,
\begin{equation}
 M'^{mass}
= U^{0T} \left[M_D^T(M_R^{flavour})^{-1}M_D \right] U^0 = 
{m_D^2 \over \sqrt 2 ~xy m_R} 
\pmatrix{0 & y & y \cr y & {x \over  \sqrt 2} & -{x \over \sqrt 2} \cr 
y & -{x \over \sqrt 2} & {x \over \sqrt 2}}\;\;.
\label{pert1}
\end{equation}
The changes in the solar sector are determined by the $2\times2$
submatrix   of $M'^{mass}$, 
\begin{equation}
M'^{mass}_{2\times2} = {m_D^2 \over \sqrt 2 ~xy m_R} 
\pmatrix{0 & y \cr y & {x/\sqrt 2}} \;.
\label{solr}
\end{equation}
From the above one has
\begin{equation}
\tan 2\theta_{12}= 2 \sqrt 2 \left(\frac{y}{x}\right)  \; .
\label{solangr}
\end{equation}
The tribimaximal mixing value of $\theta_{12}$,  which is
disallowed by the $1\sigma$ data but is allowed in the 3$\sigma$
range\footnote{From \cite{Gonzalez} we use $7.03  \leq \Delta
m_{21}^2/ 10^{-5} \, {\rm eV}^2 \leq 8.03 \;\; {\rm and}\;\;
31.30^\circ \leq \theta_{12} \leq 35.90^\circ$ at 3$\sigma$.}, is
obtained if $y/x = 1$.  From the data,  $\tan 2\theta_{12} > 0$
always, forcing $x$ and $y$ to have the same sign.  Thus it can
be inferred that either $\phi_1 = 0 = \phi_2$ or $\phi_1 = \pi =
\phi_2$. The global fits of $\theta_{12}$ provide a bound on this
ratio as,
\begin{equation}
0.682 < \frac{y}{x} < 1.075 ~{\rm at} ~3\sigma \;.
\label{t12lim}
\end{equation}
From eq. (\ref{solr}),
\begin{equation}
\Delta m^2_{solar}=  {m_D^2 \over xy m_R} ~m^{(0)}_1
\sqrt{x^2 + 8y^2}\;. 
\label{solspltr}
\end{equation} 
The first order corrected third wave function $|\psi_3\rangle$ is:
\begin{equation}
|\psi_3\rangle =
\pmatrix{\kappa \cr {1\over \sqrt 2}{(1 - \frac{\kappa}{\sqrt 2} {x\over y})} 
\cr {1\over \sqrt 2}{(1 + \frac{\kappa}{\sqrt 2} {x\over y})} } \ \ \ ,
\label{psi3_1}
\end{equation} 
where
\begin{equation}
\kappa \equiv {m_D^2 \over {\sqrt 2 ~x m_R m^-}} \;\;.
\label{kappa}
\end{equation} 
If $x > 0$  the sign of $m^-$ determines that of $\kappa$.
Comparing eq.  (\ref{psi3_1}) with the third column of the PMNS
matrix, we write,
\begin{equation}
\sin \theta_{13}\cos\delta=\kappa={m_D^2 \over {\sqrt 2 ~x m_R m^-}}\;\;,
\label{s13}
\end{equation}

For normal mass ordering (NO), $\delta = 0$ while for inverted mass ordering
(IO) $\delta = \pi$ if $x > 0$, both being CP conserving\footnote{The usual
 convention of all the mixing angles $\theta_{ij}$ belonging
 to the the first quadrant is followed.}.  
For $x < 0$ NO (IO) corresponds to $\delta = \pi ~(0)$.
From eqs. (\ref{s13}), (\ref{solangr}), and
(\ref{solspltr}) we get,
\begin{equation}
\Delta m^2_{solar} =  {\rm sgn}(x)~m^- m^{(0)}_1 
~\frac{4 \sin \theta_{13} \cos\delta}{\sin 2\theta_{12}}  \;, 
\label{solsplr2}
\end{equation}
which relates the solar sector with $\theta_{13}$. The
requirement $\Delta m^2_{solar} > 0$ is ensured by ${\rm sgn}(x)~m^- \sin
\theta_{13} \cos\delta > 0$  from eq. (\ref{s13}).
If the neutrino mass splittings, $\theta_{12}$,  
and $\theta_{13}$ are given, eq.
(\ref{solsplr2}) determines the lightest neutrino mass, $m_0$.

Inverted ordering is excluded by eq. (\ref{solsplr2}) as we now show.
If $z \equiv m^- m^{(0)}_1/\Delta m^2_{atm}$ and
$ m_0/\sqrt{|\Delta m^2_{atm}|} \equiv \tan \xi$, then
\begin{eqnarray}
z &=& \sin \xi/(1+ \sin \xi) \;\; 
{\rm (normal ~ordering)},\nonumber \\
z &=& 1/(1+ \sin \xi) \;\; {\rm (inverted  ~ordering)} \;\;. 
\label{m_0}
\end{eqnarray} 
It is seen  that $0 \leq z \leq 1/2$ for NO
and $1/2 \leq z \leq 1$ for IO and as $z \rightarrow 1/2$ 
one approaches  quasidegeneracy,
i.e., $m_0 \rightarrow $ large, in both
cases.  From eq. (\ref{solsplr2}) 
\begin{equation}
z = \left(\frac{\Delta m^2_{solar}}{|\Delta m^2_{atm}|}\right) 
\left(\frac{\sin 2\theta_{12}}{4 \sin \theta_{13} |\cos\delta|}\right) \;\;,
\label{z}
\end{equation} 
where $|\cos\delta| = 1$ for real $M_R$.  For the observed ranges of
the oscillation parameters   $z\sim 10^{-2}$, as a result
of which inverted mass ordering is disallowed.

From eq. (\ref{psi3_1}): 
\begin{equation}
\tan\theta_{23} \equiv \tan (\pi/4 - \omega) = 
\frac{1-\frac{\kappa}{\sqrt 2} {x\over y}}  
{1 + \frac{\kappa}{\sqrt 2} {x\over y}}   
\ \ , 
\label{th23r}
\end{equation}
Using eqs. (\ref{solangr}) and (\ref{s13}) in the above we get,
\begin{equation}
\tan \omega = 
\frac{2 \sin \theta_{13}\cos\delta}{\tan2\theta_{12}} \;.
\label{phir}
\end{equation}
The octant of $\theta_{23}$ is dictated by the sign of  $\omega$
which in its turn is determined by $\delta$.
$\theta_{23}$ lies in the first (second) octant, when 
$\omega$ is
positive (negative), i.e., $\delta = 0 ~(\pi)$.  
For NO, the only allowed option, this corresponds to $x > 0$ ($x < 0$).

\begin{figure}[h]
\begin{center}
\includegraphics[width=3.2in]{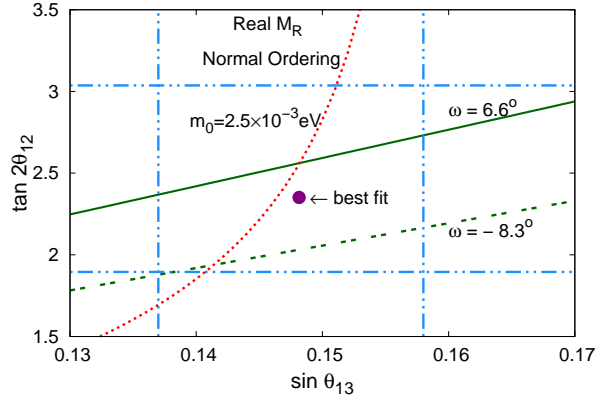}
\end{center}
\caption{ \sf  \small The  3$\sigma$  range of $\sin \theta_{13}$
and $\tan 2\theta_{12}$ from global fits is represented by the
blue dot-dashed box with the best-fit shown as a violet dot.
When the best-fit values of the two mass-splittings are used, eq.
(\ref{solsplr2}) gives  
the red dotted curve  for $m_0 = 2.5$
meV. From eq. (\ref{phir}) for
the first (second) octant 
the portion below the green solid (dashed) straight line  is
excluded by $\theta_{23}$ at 3$\sigma$. Eq.  (\ref{solsplr2})
does not allow inverted ordering  for real $M_R$. }
\label{Real1} 
\end{figure}


Our results for the real perturbation are shown in Fig.
\ref{Real1}. The 3$\sigma$ global-fit range of $\sin
\theta_{13}$ and $\tan 2\theta_{12}$ is marked by the blue
dot-dashed box and the best-fit value is indicated by a violet
dot in Fig \ref{Real1}. For any point in this region, if the two mass
splittings are specified, the $z$ (or equivalently $m_0$)
 that produces the correct solar splitting is
determined by eq. (\ref{z}).

From the 3$\sigma$ data $\omega_{min} = 0$ for both octants and 
$\omega_{max} = 6.6^\circ ~(-8.3^\circ)$ for the first
(second) octant \cite{Gonzalez}.  In this model in case of real
$M_R$ we get $|\omega| \geq 5.14^\circ$ for both octants using eq.
(\ref{phir}), as $|\cos\delta | = 1$.  This limits the range in
which $\theta_{23}$ can be obtained\footnote{This range is
excluded at 1$\sigma$ for the first octant.}. Eq. (\ref{phir})
for $\omega_{max}$ for the first (second) octant is denoted by
the green solid (dashed)  straight lines below which the model
does not hold in each case. Needless to mention that the best-fit
point  is allowed only if $\theta_{23}$ is in the second octant.

One obtains $z_{max} = 6.03 \times 10^{-2}$ using the 3$\sigma$
limits of $\theta_{13}$ and $\theta_{12}$ in eq. (\ref{z}),
implying  $(m_0)_{max}$ = 3.10 meV.  The consistency of eq.
(\ref{z}) with eq.  (\ref{phir}) at $\omega_{max}$ sets $z_{min}$
= 4.01 $\times 10^{-2}$ (3.88 $\times 10^{-2}$) for the first
(second) octant corresponding to $(m_0)_{min}$ = 2.13 (2.06) meV.
For example, if $m_0 = 2.5$ meV and if the best-fit values of the
solar and atmospheric mass splittings are used then eq.
(\ref{solsplr2}) yields the red dotted curve in Fig. \ref{Real1}.

The free parameters here are $m_0$, $m_D^2/xm_R$ and $y$ for real
$M_R$ with which the solar mass splitting, $\theta_{12},
\theta_{13}, \theta_{23}$ are obtained for normal mass ordering.
Inverted ordering is not allowed so long as the perturbation is
real.


\section{Complex $M_R$}
For the more general complex  $M_R$ in the mass basis one gets in
place of eq. (\ref{pert1}):
\begin{equation}
 M'^{mass} =
{m_D^2 \over \sqrt 2 x y m_R} 
\pmatrix{0 & y e^{i\phi_1} & y e^{i\phi_1} \cr y e^{i\phi_1} &
x {e^{i\phi_2}}\over{\sqrt{2}} & - x {e^{i\phi_2}}\over{\sqrt{2}} \cr
y e^{i\phi_1} & -{x e^{i\phi_2}}\over{\sqrt{2}}&
x {e^{i\phi_2}}\over{\sqrt{2}}}.
\label{TypeIcmplxM}
\end{equation}

Here $x$ and $y$ are  positive. One observes that $M'$ in
eq. (\ref{TypeIcmplxM}) is not hermitian. To proceed, one chooses
the hermitian combination $(M^0 + M')^\dagger(M^0 + M')$ treating
$M^{0\dagger} M^0$ as the zeroth  order term and $(M^{0\dagger}
M' + M'^\dagger M^0)$ as the lowest order perturbation. The
unperturbed eigenvalues now are $(m^{(0)}_i)^2$ and the
perturbation matrix, which is hermitian by construction,  is
\begin{equation}
(M^{0\dagger} M' + M'^\dagger M^0)^{mass} = 
{m_D^2 \over \sqrt 2 xy m_R}
\pmatrix{ 0 & 2 m^{(0)}_1 y \cos\phi_1 & 
y f(\phi_1) \cr
2 m^{(0)}_1 y \cos\phi_1 & { 2 \over \sqrt{2}}m^{(0)}_1 x \cos\phi_2 & 
-{1\over\sqrt{2}} x f(\phi_2)\cr
y f^*(\phi_1)& 
 -{1\over\sqrt{2}} x f^*(\phi_2)& 
{ 2 \over \sqrt{2}}m^{(0)}_3 x \cos\phi_2},
\label{pertcmplx}
\end{equation}
with
\begin{equation}
f(\xi) = m^{+} \cos\xi - i m^{-} \sin\xi \;\;.
\label{ffn}
\end{equation}
Beyond this point steps similar to those for real $M_R$ are followed.

From eq. (\ref{pertcmplx}) the  solar mixing angle now is
\begin{equation}
\tan 2\theta_{12}= 2\sqrt2 ~{y \over x}
~{\cos\phi_1\over\cos\phi_2} \;.
\label{solangcmplx}
\end{equation}
Thus, $(\cos\phi_1/\cos\phi_2)$ has to be positive. The limits
given in eq. (\ref{t12lim}) will now apply on the combination
$(y/x)(\cos\phi_1/\cos\phi_2)$.   

In the complex $M_R$ case including first order corrections 
$|\psi_3\rangle$  becomes 
\begin{equation}
|\psi_3\rangle =
\pmatrix{\kappa f(\phi_1)/m^+\cr {1\over \sqrt 2}{(1-{\kappa \over \sqrt
2}\frac{x}{y} ~f(\phi_2)/m^+)} \cr {1\over \sqrt 2}{(1+{\kappa\over \sqrt
2}  \frac{x}{y} ~f(\phi_2)/m^+)}
} .
\label{psi3ca}
\end{equation}
Now $\kappa$ is positive (negative) for NO (IO) always.
Eq. (\ref{psi3ca}) implies
\begin{eqnarray}
\sin \theta_{13}\cos\delta &=& \kappa  \cos\phi_1 \ , 
\nonumber \\ 
\sin \theta_{13}\sin\delta &=& \kappa ~\frac{m^-}{m^+} \sin\phi_1\  \ . 
\label{s13cmplx}
\end{eqnarray}
So, $\cos\delta$ has the same sign  (opposite sign) as that of $\cos
\phi_1$ for NO (IO). Further,  the
product $\sin \theta_{13} \sin\delta$  in the CP-violation
Jarlskog parameter, $J$, is dependent on $\sin \phi_1$.  The phase
$\phi_2$ has no affect on $\delta$.

For normal ordering -- $\kappa > 0$ --  the quadrants of $\delta$
and $\phi_1$ are the same while for inverted ordering -- $\kappa
< 0$ -- $\delta$ has to be in the first (third) quadrant when
$\phi_1$ happens to be in the second (fourth) quadrant and {\em
vice-versa}. So, a near-maximal $\delta = 3\pi/2 - \epsilon$ can
be obtained if $\phi_1 \sim 3\pi/2 - \epsilon ~(3\pi/2 + \epsilon)$
for normal (inverted) ordering.

From eq. (\ref{psi3ca}) 
\begin{equation}
\tan\theta_{23} = {{1-{\kappa \over
\sqrt 2}\frac{x}{y}\cos\phi_2} \over  {1+{\kappa\over \sqrt
2}\frac{x}{y} \cos\phi_2}}  \ \ \ \ ,
\end{equation}
Using eqs. (\ref{solangcmplx}) and (\ref{s13cmplx}),
\begin{equation}
\tan\omega 
= \frac{2\sin \theta_{13}\cos\delta}{\tan2\theta_{12}} \;.
\label{phic}
\end{equation}
The corresponding result for real $M_R$ -- eq. (\ref{phir}) -- is
recovered if $\cos \delta = \pm 1$. From eq. (\ref{phic})
$\theta_{23}$ is in the first octant if $\delta$ lies in the
first or the fourth quadrant -- which result in opposite signs of $J$
-- otherwise it is in the second octant. This correlation does
not depend on the mass ordering. 
Thus the first (second) octant goes with $\delta =
3\pi/2 + \epsilon ~(3\pi/2 - \epsilon)$ if $\delta$ is near $3 \pi/2$.

If $m_D$ and $m_R$ are expressed in terms of
$\sin \theta_{13}\cos\delta$, one finds
\begin{equation}
\Delta m^2_{solar}
=  {\rm sgn}(\cos\phi_2) ~m^- m^{(0)}_1 
~\frac{4 \sin \theta_{13} \cos\delta }{\sin 2\theta_{12}} 
\;,
\label{solsplc}
\end{equation}
which is of very similar form as  eq. (\ref{solsplr2}) for real
$M_R$. Obviously, eqs. (\ref{m_0}) and (\ref{z}) still apply.  If
one notes the factors which determine  the sign of $\cos\delta$
one can conclude that the positivity of $\Delta m^2_{solar}$ is
ensured if sgn$(\cos\phi_1
\cos\phi_2)$ is positive for both mass orderings. Therefore, the
sign of the solar mass splitting accommodates both octants of
$\theta_{23}$ irrespective of the mass ordering.
The admissible range of $\delta$ can be identified by
reexpressing eq. (\ref{z}) as:
\begin{equation}
|\cos\delta| = \left(\frac{\Delta m^2_{solar}}{|\Delta m^2_{atm}|}\right) 
\left(\frac{\sin 2\theta_{12}}{4 \sin \theta_{13} ~z}\right) \;\;.
\label{cdel}
\end{equation}


\begin{figure}[h]%
\begin{center}
  \parbox{2.45in}{\includegraphics[width=2.59in]{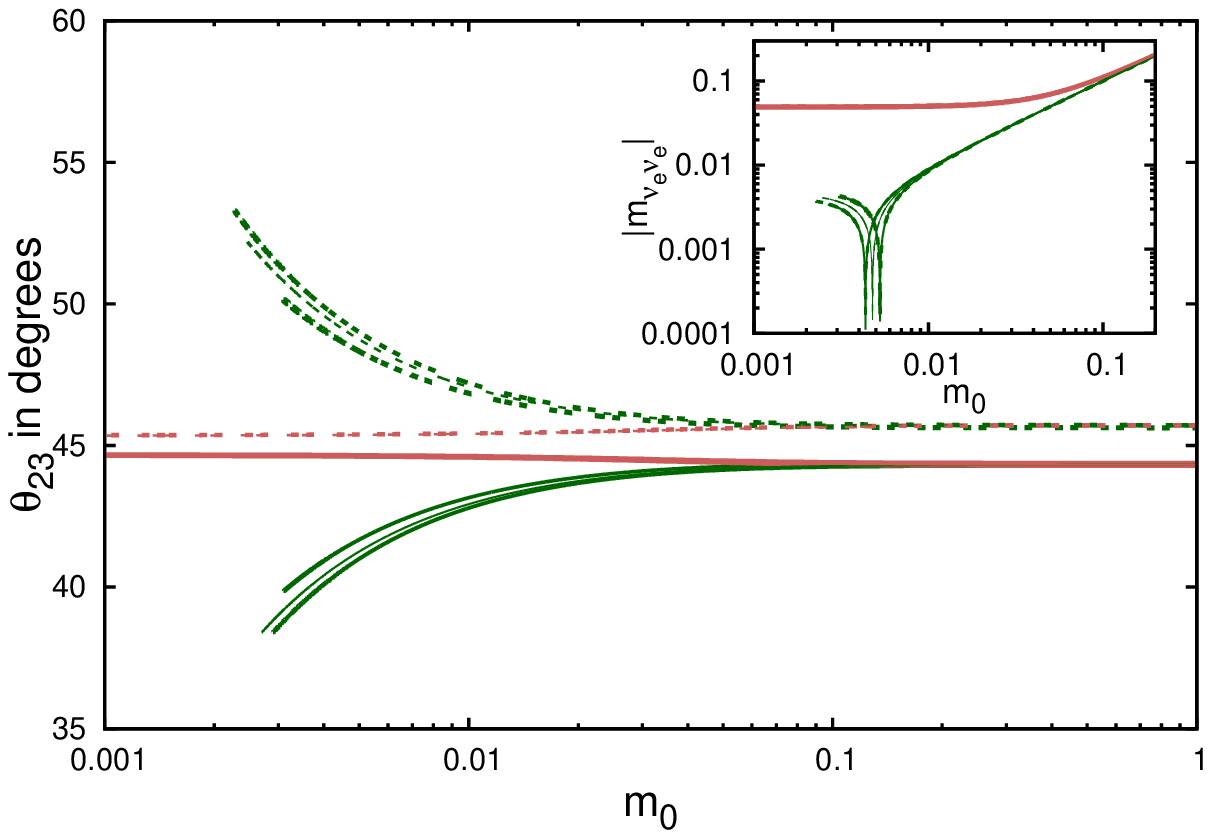}}
  \hspace*{1.28pt}
  \parbox{2.45in}{\includegraphics[width=2.59in]{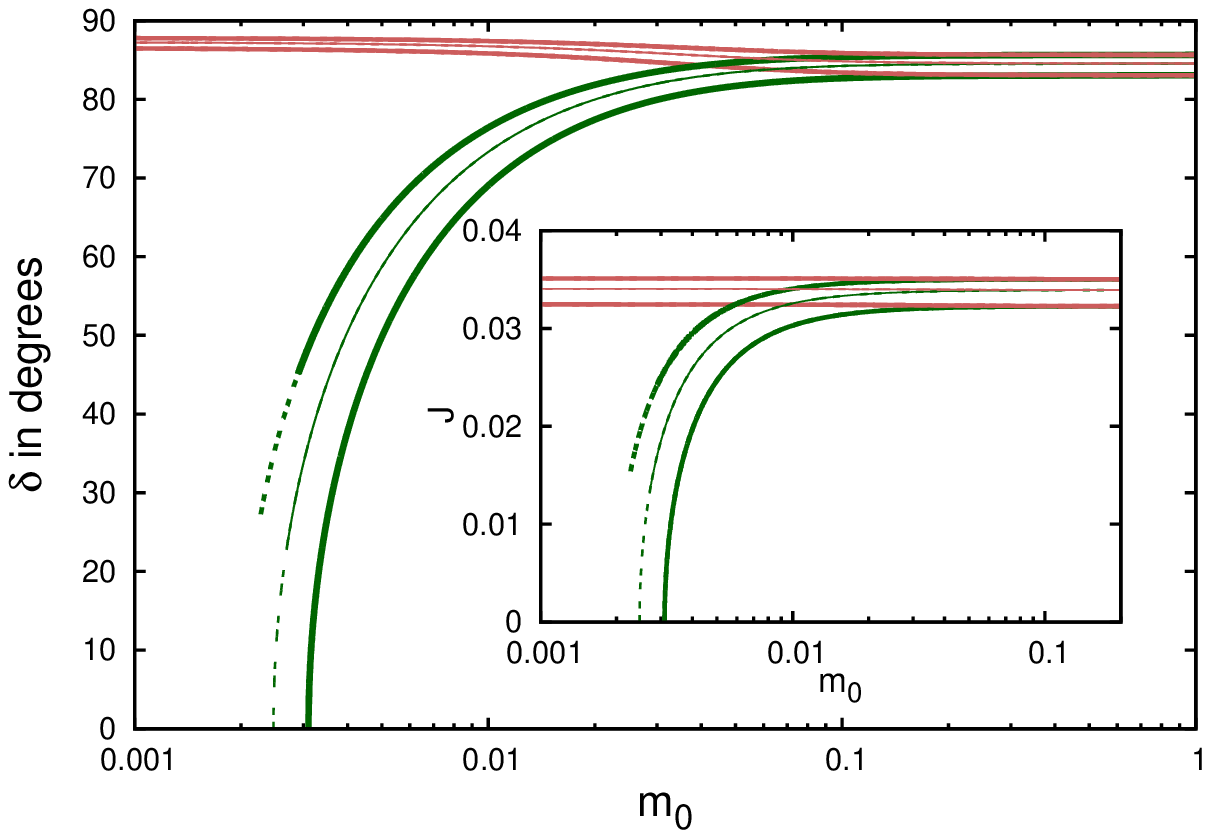}} 
\caption{\sf \small  $\theta_{23}$, $|m_{\nu_e \nu_e}|$  (in eV),
$\delta$, and  $J$ as a function of the lightest neutrino mass
$m_0$ (in  eV).  The green (pink) curves are for NO (IO). The
3$\sigma$ allowed region is between the thick curves of each type
while the
thin curves are for the best-fit input values. The solid (dashed)
curves are for the first (second) octant of $\theta_{23}$.  Left:
The variation of $\theta_{23}$. The inset shows $|m_{\nu_e
\nu_e}|$  (in  eV),  the effective mass controlling $0\nu2\beta$
processes.  Right: CP-phase $\delta$. The inset  exhibits the
CP-violation measure $J$.  }
\label{CP} 
\end{center} 
\end{figure} 

In the analysis which we report $m_0,  \theta_{13}$, and
$\theta_{12}$ are the inputs. We get $\delta$ and $\theta_{23}$
from eqs. (\ref{cdel}) and (\ref{phic}).  The CP-violation
measure, $J$, and the combination $|m_{\nu_e \nu_e}|$ which
contributes to $0\nu 2\beta$  are then easily obtained.

The results for complex $M_R$ are presented in Fig. \ref{CP}. The
left panel (thick curves) shows the variation of $\theta_{23}$
with  $m_0$ when the neutrino mass square splittings  and the
angles $\theta_{13}$ and $\theta_{12}$ cover their 3$\sigma$
ranges. The thin curves are for the best-fit values.  In this
figure the green (pink) curves are always for NO (IO) while solid
(dashed) curves are for $\theta_{23}$ in the first (second)
octant.  For IO the thick and thin curves are too close for
distinction in this panel. Note that $\theta_{23} = \pi/4$ is not
consistent with the 3$\sigma$ predictions from this model. It is
seen that $\theta_{23}$ is symmetrically distributed about
$\pi/4$, which is expected from eq.  (\ref{phic}). For IO the
obtained range  is outside 1$\sigma$ but are admissible at
3$\sigma$. When $\theta_{23}$ is better measured  one of the mass
orderings will be eliminated unless the neutrinos are in the
quasidegenerate regime.

The 3$\sigma$ limits of $\theta_{23}$ in the two octants
determine the minimum permitted value of $m_0$ for NO.  For IO
eq. (\ref{solsplc}) allows $m_0$ to be arbitrarily small (see
below).  In the inset of this panel $|m_{\nu_e \nu_e}|$ has been
plotted.  Direct neutrino mass measurements \cite{katrin} are
expected to be sensitive to masses up to 200 meV.  Planned $0\nu
2\beta$  experiments will access $m_0$ in  the quasidegenerate
regime \cite{0nubeta}.  Fig. \ref{CP} indicates that to
distinguish the alternate mass ordering possibilities an order of
magnitude improvement in their sensitivity will be needed. Large
atmospheric neutrino detectors such as INO or long-baseline
experiments are alternate avenues for determining the mass
ordering.

In  the right panel of Fig. \ref{CP} the variation of $\delta$
with $m_0$  for both mass orderings is shown. The dependence of
$J$ appears in the inset.  The two panels of  Fig. \ref{CP} use
the same conventions.  Since the three mixing
angles are kept in the first quadrant, $J$ is positive if $0 \leq
\delta \leq \pi$ and is negative otherwise.  As mentioned before,
the quadrant of $\delta$  can be altered by choosing the quadrant
of $\phi_1$ suitably\footnote{From eq. (\ref{s13cmplx}), $\delta
\rightarrow \pi + \delta$ if $\phi_1 \rightarrow \pi + \phi_1$.}.
However,  for a particular mass ordering from eq.  (\ref{cdel})
the dependence of $|\cos\delta|$ on $m_0$ is the same for the
different alternatives, which are $\pm
\delta$ and $(\pi \pm \delta)$.  Keeping  this  in mind, in Fig.
\ref{CP} (right panel) $\delta$  has been plotted in the
first quadrant and $J$ has been taken as positive.

As $\theta_{23}$ is symmetric around $\pi/4$ in this model and
$J$ is proportional to $\sin 2\theta_{23}$ so it is independent
of the  octant.   For inverted mass ordering both $\delta$ and
$J$ remain nearly unaffected by variations of $m_0$.

If $m_0$ is smaller than 10 meV, then the CP-phase $\delta$ is
much larger for inverted ordering. Once the mass ordering is
known and CP-violation in the neutrino sector is measured this
could provide a clear test of this model.  Consistent with Sec.
\ref{sec3}, the limit of real $M_R$ is admissible  only for NO,
and that too for only a portion of the 3$\sigma$ range.

As discussed, one has  $0 \leq z \leq  1/2$ for NO and $1/2 \leq
z \leq 1$ for IO. It is  seen from eq. (\ref{cdel}) that as a
consequence of this the
allowed $\delta$  are complementary for the two mass orderings
tending towards a common value in the
quasidegenerate limit, which sets in from around $m_0$ =
100 meV.  Unlike the real $M_R$ case, in eq.
(\ref{z}) by taking $\cos\delta$  small one can
make $z \equiv m^- m^{(0)}_1/\Delta m^2_{atm} \sim 1$ so that
solutions exist for $m_0$ for IO 
corresponding to even $m_0$ arbitrarily small unlike for NO 
where the lower limit of $m_0$ is set by $\cos\delta =
1$, i.e., real $M_R$.  




\section{Conclusions}

Summarizing, a neutrino mass model is presented in which the
observed solar mass splitting, $\theta_{12}$, $\theta_{13}$, and
$\omega = \pi/4 - \theta_{23}$ all originated from a single
perturbation (derived out of a Type-I see-saw mechanism) and are
thereby related to each other. The   atmospheric
mass splitting preferred by the data and  maximal mixing in this
sector play the role  of the unperturbed framework. In order
to restrict free parameters to a minimum the
Dirac term in the see-saw is taken as proportional to the
identity matrix and the right-handed neutrino mass matrix, $M_R$,
has a four-zero texture in the flavour basis. Requiring
that the mixing angles and solar mass splitting identified by the
global fits be reproduced, for a real $M_R$ a narrow range
of the lightest neutrino mass ($m_0 \sim $ a few meV) is
permitted for normal ordering. It leaves the option
open for $\theta_{23}$ to belong to the first or the second
octant.  Such a CP conserving real perturbation forbids inverted
ordering.  The more general complex $M_R$ enables a
{considerable enlargement of the range of $m_0$ and determines in its
terms the CP-phase $\delta$ and the octant of $\theta_{23}$} as
well, while accommodating both mass orderings. In the
quasi-degenerate limit and in case of inverted ordering $\delta
\sim 3\pi/2$ is a natural prediction.  Future improved
measurements of $\delta$, $\theta_{23}$, $0\nu 2\beta$
and determination  of the neutrino mass ordering will
test the model from various angles.

\section*{Acknowledgements} AR thanks the organizers for
arranging a very stimulating meeting on neutrino physics. SP
acknowledges a Senior Research Fellowship from CSIR, India.  AR
is partially funded by  the Department of Science and Technology
Grant No. SR/S2/JCB-14/2009.



\end{document}